%% file: PaperForReview.tex
\documentclass[10pt,twocolumn,letterpaper]{article}

\usepackage[pagenumbers]{cvpr} 

\usepackage{graphicx}
\usepackage{amsmath}
\usepackage{amssymb}
\usepackage{booktabs}
\usepackage[table,xcdraw]{xcolor}

\usepackage[pagebackref,breaklinks,colorlinks]{hyperref}
\usepackage{multirow}
\usepackage[accsupp]{axessibility}

\usepackage[capitalize]{cleveref}
\crefname{section}{Sec.}{Secs.}
\Crefname{section}{Section}{Sections}
\Crefname{table}{Table}{Tables}
\crefname{table}{Tab.}{Tabs.}

\def\cvprPaperID{5284} 
\def\confName{CVPR}
\def\confYear{2023}

\begin{document}

\title{Fully Self-Supervised Depth Estimation from Defocus Clue}


\author{
    Haozhe Si\textsuperscript{1,2}\thanks{\indent The contributions by Haozhe Si have been conducted and completed during his internship at Shanghai AI Laboratory.} \thanks{\indent Equal contribution.}  \hspace{0.5cm}
    Bin Zhao\textsuperscript{1,3}\footnotemark[2] \hspace{0.5cm}
    Dong Wang\textsuperscript{1}\thanks{\indent Corresponding author.}\hspace{0.5cm}
    Yunpeng Gao\textsuperscript{3} \hspace{0.5cm} \\
    Mulin Chen\textsuperscript{1,3} \hspace{0.5cm}
    Zhigang Wang\textsuperscript{1} \hspace{0.5cm}
    Xuelong Li\textsuperscript{1,3}\footnotemark[3] \vspace{0.5cm} \\ 
    \textsuperscript{1}Shanghai AI Laboratory \hspace{0.5cm}
    \textsuperscript{2}University of Illinois Urbana-Champaign \\
    \textsuperscript{3}Northwestern Polytechnical University \\
    {\tt\small \href{mailto:haozhes3@illinois.edu}{haozhes3@illinois.edu}, \{\href{mailto:zhaobin@pjlab.org.cn}{zhaobin}, \href{mailto:wangdong@pjlab.org.cn}{wangdong}\}@pjlab.org.cn, \href{mailto:li@nwpu.edu.cn}{li@nwpu.edu.cn}}
}
\maketitle

\begin{abstract}
  Depth-from-defocus (DFD), modeling the relationship between depth and defocus pattern in images, has demonstrated promising performance in depth estimation. Recently, several self-supervised works try to overcome the difficulties in acquiring accurate depth ground-truth. However, they depend on the all-in-focus (AIF) images, which cannot be captured in real-world scenarios. Such limitation discourages the applications of DFD methods. To tackle this issue, we propose a completely self-supervised framework that estimates depth purely from a sparse focal stack. We show that our framework circumvents the needs for the depth and AIF image ground-truth, and receives superior predictions, thus closing the gap between the theoretical success of DFD works and their applications in the real world. In particular, we propose (i) a more realistic setting for DFD tasks, where no depth or AIF image ground-truth is available; (ii) a novel self-supervision framework that provides reliable predictions of depth and AIF image under the challenging  setting. The proposed framework uses a neural model to predict the depth and AIF image, and utilizes an optical model to validate and refine the prediction. We verify our framework on three benchmark datasets with rendered focal stacks and real focal stacks. Qualitative and quantitative evaluations show that our method provides a strong baseline for self-supervised DFD tasks. The source code is publicly available at \texttt{\href{https://github.com/Ehzoahis/DEReD}{https://github.com/Ehzoahis/DEReD}}.
\end{abstract}

\section{Introduction} \label{sec:intro}
Depth estimation is a fundamental task in computer vision. Predicting depth from RGB images shows great potential in applications like autonomous cars or robots. It has also been used as a powerful pretext training task in unsupervised downstream tasks including visual feature extraction \cite{jiang2018self, ren2018cross}, semantic segmentation \cite{hoyer2021three, cardace2022plugging}, \textit{etc}. However, collecting sufficiently diverse datasets with per-pixel depth ground-truth for supervised learning is challenging. To overcome this limitation, self-supervised depth estimation works exploit the consistency of 3D geometries within stereo pairs \cite{garg2016unsupervised, godard2017unsupervised} or monocular videos \cite{ji2021monoindoor, godard2019digging}, and have shown impressive results. Nevertheless collecting synchronized multi-view images or videos is resource-consuming.

\begin{figure}[!t]
    \centering
    \includegraphics[trim=30 10 80 10,clip, width=\columnwidth]{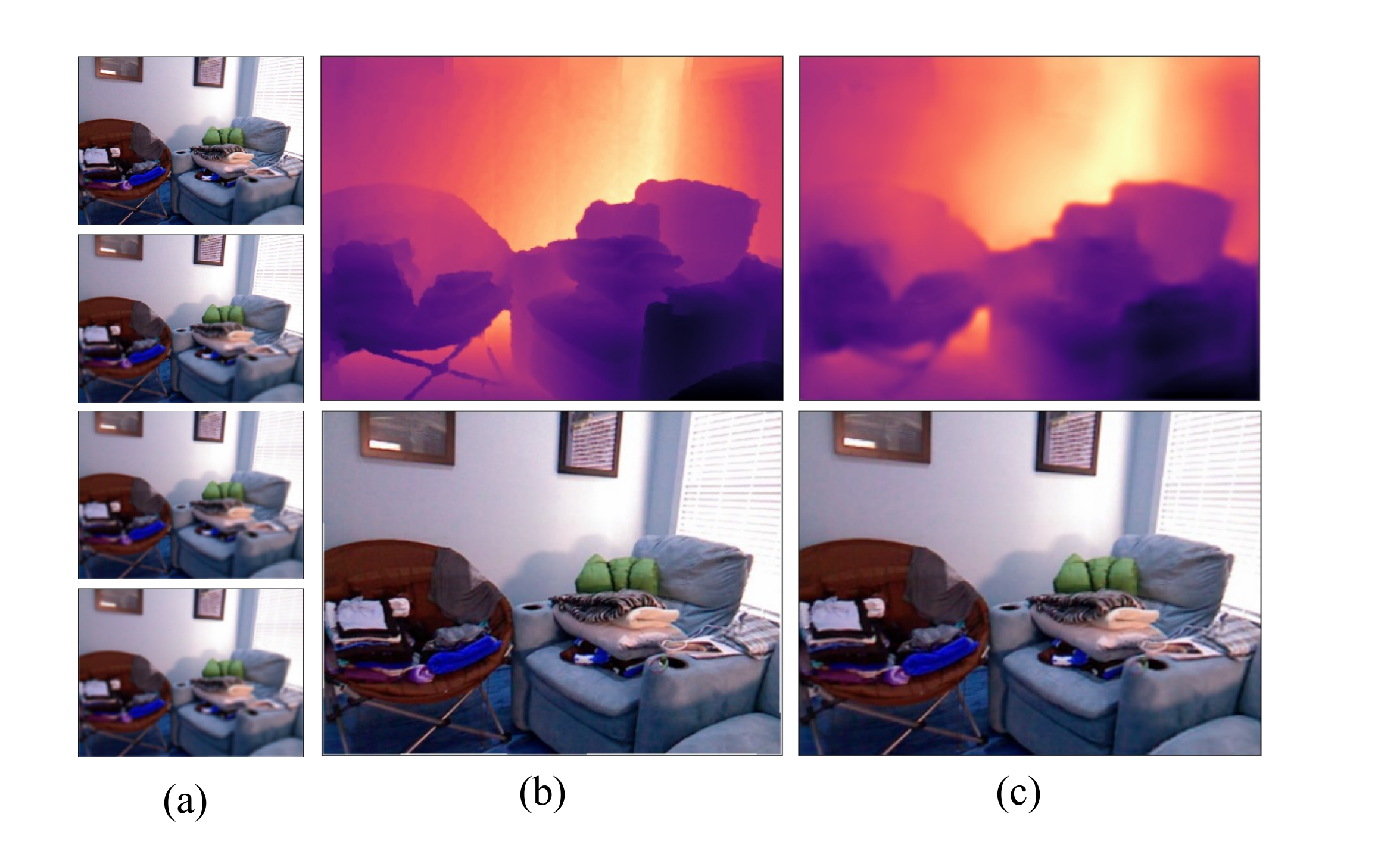}
    \caption{(a) A sparse focal stack from the NYUv2 dataset \cite{Silberman:ECCV12}. (b) The ground-truth depth map and all-in-focus (AIF) image. (c) The depth map and AIF image estimated by our self-supervised framework from the sparse focal stack.}
    \label{fig:head}
\end{figure}

Another clue for depth estimation is defocus. Defocus blur is a measurable property of images that is associated with the geometry of camera lens and the changes in depth. Previous works \cite{favaro2003observing, 7298972} prove that depth can be recovered from a set of images with different focus distances, \textit{i.e.}, a focal stack, by observing their blurry amounts. With the help of deep learning methods, supervised DFD works \cite{yang2022deep, 9157368} estimate depth in a data-driven way. Recent DFD methods \cite{8954440, Wang-ICCV-2021, 9464709} claim they are freed from being supervised by the depth ground-truth. However, these works utilize the all-in-focus (AIF) image in model training, which is more of a theoretical concept than images captured by camera lens \cite{kingslake1989history} in the real world. Existing works generally treat images taken with small apertures as the AIF image. Such approximations inevitably contain regional blurriness and suffer from underexposure in short range. Thus, current DFD methods have drawbacks in piratical applications.

In order to close the gap between theories and real-world applications, we design a more realistic setting for the DFD tasks: only focal stacks are provided in model training, while the availability of depth and AIF images ground-truth is deprived. This is a challenging task because we no longer have the direct/indirect optimization goal for the model. To tackle this challenge, we propose a self-supervised DFD framework, which constrains the predicted depth map and AIF image to be accurate by precisely reconstructing focal stacks. Specifically, our framework consists of a neural model, \textbf{D}epth\textbf{AIF}-\textbf{Net} (\textbf{DAIF-Net}), which predicts the AIF image and the depth map from the focal stack, and a phyisical-realistic optical model that reconstructs the input from the predicted AIF image and depth map. Since all images in a focal stack are sharing the same depth and AIF image, and the physics model can deterministically map them to focal stacks, accurate depth maps and AIF images are a necessary and sufficient condition for precisely reconstructing the input. Therefore, by assuring the consistency between the input and the reconstructed focal stack, the prediction of depth map and AIF image can be intermediately improved. To the best of our knowledge, this is the first self-supervised DFD work that relieves the need for AIF images and depth ground-truth. Such improvements over previous supervised and indirectly-supervised works make our method more applicable in real scenarios.


Extensive experiments are conducted on both synthetic and real datasets. Results show that the proposed framework is on par with the state-of-the-art supervised/indirectly supervised depth-from-focus/defocus methods and has convincing visual quality, as shown in \Cref{fig:head}. Additionally, we apply our model to the data in the wild, and demonstrate the ability of our framework to be applied in real scenarios. Finally, we extend the training paradigm of our framework so that our model has the ability to be transferred between focal stacks with an arbitrary number of images. 

Our contribution is three-fold:
\begin{itemize}
    \item We design a more realistic and challenging scenario for the Depth-from-Defocus tasks, where only focal stacks are available in model training and evaluation.
    \item We propose the first completely self-supervised framework for DFD tasks. The framework predicts depth and AIF images simultaneously from a focal stacks and is supervised by reconstructing the input.
    \item Our framework performs favorably against the supervised state-of-the-art methods, providing a strong baseline for future self-supervised DFD tasks.
\end{itemize}
\section{Related Works}

\subsection{Self-supervised Monocular Depth Estimation} 
Monocular depth estimation works aim to regress a dense depth map from a single RGB image. Methods for self-supervised monocular depth estimation utilize scene geometry as a constraint in training. Specifically, \cite{garg2016unsupervised} introduces the differentiable inverse warping and proposes an auto-encoder-based model trained from stereo pairs; the follow-up work \cite{godard2017unsupervised} adds left-right consistency to further exploit geometry clues. To train models from video sequences, ego-motion \cite{zhou2017unsupervised} and optical flow \cite{yin2018geonet} of the scenes are also added to the frameworks. Following these works, recent works \cite{ji2021monoindoor, bian2020unsupervised} have shown impressive results on indoor datasets. Although self-supervised monocular depth estimation works are capable of producing depth maps without training with depth ground-truth, they require a large number of stereo pairs and/or monocular video sequences. Additionally, because the absolute scales are ambiguous in monocular depth estimation, such models are facing challenges in transferring to unseen datasets. 

\subsection{Depth from Focus/Defocus} 
Depth estimation from focus/defocus works reconstructs the depth map by observing the blurriness of images caused by the lens effect. Analytical depth-from-focus (DFF) and depth-from-defocus (DFD) approaches \cite{favaro2003observing, watanabe1996real, nayar1994shape} calculate depth by measuring the sharpness of each pixel. Both approaches result in low-quality depth maps. Optimization-based DFF methods \cite{7271087, 7298972} usually take dense focal stacks as input, which produce high-quality depth maps but are time-consuming. Recent deep methods are trying to estimate depth from sparse focal stacks or a single defocus image. \cite{Carvalho2018eccv3drw} shows that images with defocus blur aid the model to learn depth better. \cite{9157368} transfers supervised models trained on synthetic datasets to real world datasets by learning to estimate the defocus map as an intermediate clue. \cite{Wang-ICCV-2021} further makes use of the implicit mutual information of depth and AIF image to finetune the model. \cite{9464709} trains a DefocusNet and a FocusNet simultaneously to perform DFD and DFF tasks and the models are supervised by their consistency. \cite{yang2022deep} add focus volume and differential focus volume to their supervised model to improve estimation quality. Note all the aforementioned models either require ground-truth or AIF image in addition to the focal stack, while neither of them is trivial to acquire in the wild. Motivated by this, we propose a fully self-supervised framework that requires focal stacks only.

\begin{figure*}
    \centering
    \includegraphics[trim=120 610 0 40,clip,width=.9\textwidth]{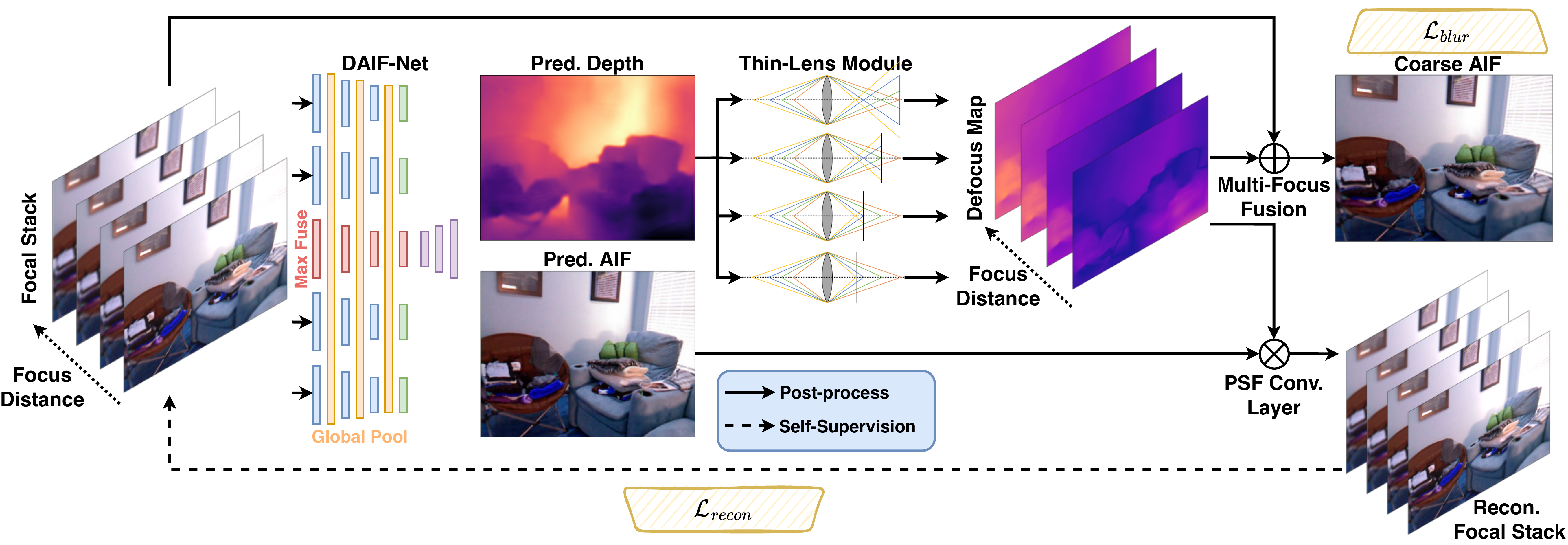}
    \caption{An overview of the proposed framework. The framework consists of a neural model, DAIF-Net, and an optical model. The optical model is composed of the thin-lens module and the PSF convolution layer. The DAIF-Net estimates the depth map and AIF image, and the optical model reconstruct the input focal stack to supervise the prediction.}
    \label{fig:ss-dfd}
\end{figure*}

\section{Proposed Method} \label{sec:method}
In this section, we introduce our framework for depth estimation from sparse focal stacks. The framework is visualized in \Cref{fig:ss-dfd}. In the training stage, the DAIF-Net estimates AIF images and depth maps from the focal stacks. The optical model then reconstructs the inputs from the predicted depth and AIF image by simulating the physical process of defocus generation. Losses and regularization are applied to encourage the detailed reconstruction, and consequently improve the quality of the estimated depth and AIF image. In the inference stage, we estimate the depth map and AIF image directly from the trained DAIF-Net without the rendering process. 

In the following, we first elaborate on the optical model behind the generation of defocus blur (\cref{sec:optical}). Then, we introduce our DAIF-Net and the intuition behind it (\cref{sec:funet}). Finally, we describe our supervision strategy that supports the self-supervised learning framework (\cref{sec:self_supervise}).

\subsection{Optical Model of Defocus Blur Generation} \label{sec:optical}
\subsubsection{Defocus Map Generation} \label{sec:defocus}

\begin{figure}
    \centering
    \includegraphics[width=\columnwidth]{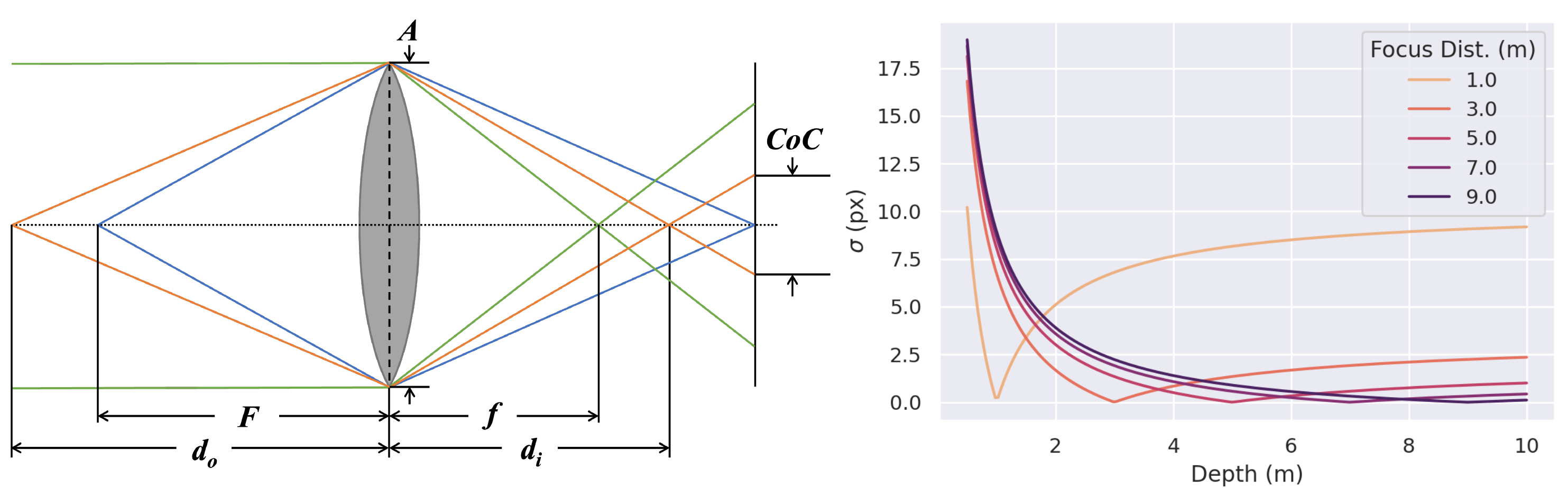}
    \caption{\textbf{Left}: Illustration of the thin-lens equation. See text for symbol definitions. \textbf{Right}: The response curve of the $CoC$ radius, $\sigma$, at different scene depth, with different focus distance.}
    \label{fig:defocus}
\end{figure}

Defocus blur is a well-studied phenomenon that naturally exists in optical imaging systems. To quantitatively measure the defocus blur in an image, we introduce the defocus map. Given an optical system, the defocus map can be calculated from the depth map once we establish the relationship between depth and defocus. As illustrated in the left of Figure \ref{fig:defocus}, when a point light source is out-of-focus, the light rays will converge either in front of or behind the image plane, and form a blurry circle on the image plane. The circle of confusion ($CoC$) measures the diameter of such a blurry circle. If the point light source is in focus, it will form an infinitely small point on the image plane, making it the sharpest projection with the minimum $CoC$. Therefore, $CoC$ describes the level of blurriness, in another word, the amount of defocus. Deriving from the thin-lens equation, the relationship between $CoC$, scene depth and camera parameters are well established:
\begin{equation}
    \centering
    CoC = A\frac{|d_o-F|}{d_o}\frac{f}{F-f},
\end{equation}
where $A$ is the aperture diameter, $d_o$ is the object distance, or depth, $F$ is the focus distance, $f$ is the focal length. In practice, the f-number, $N=f/A$, is commonly used to describe the aperture size. Since all of the parameters are in units of meters, we also need to convert $CoC$ into units of pixels. Finally, we need the radius, $\sigma$, of the circle of confusion in focal stack rendering. The resulting equation becomes:
\begin{equation} \label{eq:thin_lens}
    \sigma = \frac{CoC}{2\cdot p} = \frac{1}{2p}\frac{|d_o-F|}{d_o}\frac{f^2}{N(F-f)},
\end{equation}
where $p$ is the CMOS pixel size. With \cref{eq:thin_lens} and a set of camera parameters, we can easily convert a depth map into defocus map. By varying the focus distance $F$, we can produce the defocus map for a focal stack. The right of Figure \ref{fig:defocus} shows how defocus responds to the changing of depth at the different focus distances. Note that with the larger focus distance, it is generally harder to distinguish the amount of defocus, especially when two focus distances are close. Since training our framework heavily relies on reconstructing the focal stack, indistinguishable defocus defects the effectiveness for our method. Therefore, we tend to avoid this issue by properly selecting the camera parameters, \textit{i.e.}, the f-number $N$, the focal length $f$ and the focus distance $F$.

\subsubsection{Defocus Image Rendering} \label{sec:focal_stack}
Given the defocus map and the AIF image, we can explicitly model the generation process of the defocus image. Taking advantage of the deterministic relationship, our predicted depth and AIF image can be supervised by reconstructing the input focal stack. To render a defocus image, we convolve the AIF image with the point spread function (PSF). PSF describes the pattern of a point light source transmitting to the image plane through the camera lens. In practice, we calculate the defocus blur using a simplified disc-shaped PSF, $i.e.$, a Gaussian kernel, following previous works \cite{8099990, 8954440}. We also consider the radius of the pixel-level circle of confusion $\sigma$ as the standard deviation for the Gaussian kernel. Given the defocus map $\Sigma$ calculated in Sec. \ref{sec:defocus}, the PSF center at $x$, $y$ with $\sigma = \Sigma_{x, y}$ can be written as:
\begin{equation}
    \mathcal{F}_{x,y}(u, v) = \frac{1}{2\pi\Sigma_{x, y}^2}\exp\Big(-\frac{u^2+v^2}{2\Sigma_{x, y}^2}\Big),
\end{equation}
where $u$, $v$ is the location offset to the kernel center. Let $I$ be the AIF image and $J$ be the rendered defocus image, we produce $J$ with defocus map $\Sigma$ by convolving $\mathcal{F}$ with $I$:
\begin{equation}
    J:=I\otimes \mathcal{F}.
\end{equation}
Note $\mathcal{F}$ is a pixel-specific Gaussian kernel, therefore, unlike the traditional convolution, we need to change the kernel as the convolution window moves. Here, we adopt the PSF convolution layer \cite{8954440}. The PSF convolution layer is implemented in CUDA, thus supporting backpropagation and can be incorporated into our training pipeline. By providing the AIF image $I$ and the defocus map $\Sigma$, the PSF convolution layer will calculate the Gaussian kernel with the standard deviation clue from $\Sigma$ on the air and convolve with $I$. In practice, the window size for convolution is $7\times7$ and pixels with defocus value $\sigma<1$ will be treated as a clear pixel since its $CoC$ radius is less than one pixel. By generating multiple defocus maps and rendering the corresponding defocus images, we can construct the focal stack with varying focus distances. 

\textbf{Discussion on the Optical Model.} We choose the physics-realistic optical model to reconstruct the focal stack over a neural model mainly for two reasons. First, the optical model is a physically explainable process, therefore, an implicit constraint on the physical meanings will be applied to the inputs. In another word, the optical model encourages the predicted depth and AIF images to have their corresponding physical properties so that the focal stack can be properly reconstructed. However, using a neural model does not have such constraints. Second, it is hard to optimize the DAIF-Net model and a neural model for reconstruction jointly. In contrast, using an optical model requires no training. Meanwhile, the physics model is naturally more robust and has a provable performance.

\begin{figure}
    \centering
    \includegraphics[trim=10 443 405 0,clip,width=\columnwidth]{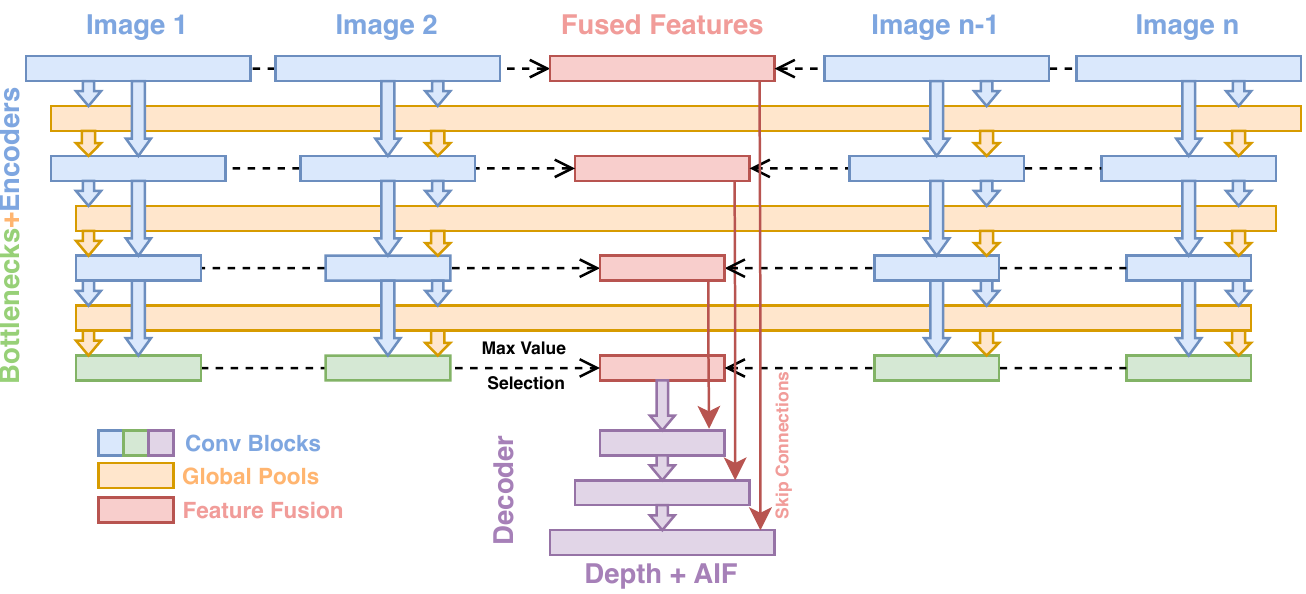}
    \caption{The DAIF-Net architecture. The architecture takes a focal stack of arbitrary size and estimates the depth map and AIF image. The parameters of encoders and bottlenecks are shared across all branches. We adopt the global pooling \cite{9157368} and fuse the branches by selecting the maxima of their features.}
    \label{fig:funet}
\end{figure}

\subsection{DAIF-Net} \label{sec:funet}
The optical model depicts the forward process of generating defocus image from the AIF image and the depth. The model calculates the following equation explicitly:
\begin{equation} \label{eq:forward}
    J_F = \mathcal{G}_F(I, D),
\end{equation} 
where $J_F$ is the defocus image with focus distance $F$, $\mathcal{G}$ is the optical model, $I$ and $D$ are the AIF image and the depth map respectively.

In contrast, predicting depth and AIF image from the focal stack, the DAIF-Net is implicitly learning the reverse expression of \cref{eq:forward}. To be specific, we are learning:
\begin{equation}
    \{I, D\} = \mathcal{D}(J_F),
\end{equation} 
where $\mathcal{D}$ is the DAIF-Net. Solving this equation using a single defocus image is an ill-posed problem because we are calculating two variables from one input. Taking advantage that the defocus images in a focal stack are sharing the same depth and AIF image. Thus, we can train our model on a focal stack to make this task solvable, and our DAIF-Net has the following expression: 
\begin{equation} \label{eq:reverse}
    \{I, D\} = \mathcal{D}(J_F^0, J_F^1, \cdots, J_F^k),
\end{equation} 
where that the accurate depth map and AIF image is the only solution to this function.

Consequently, from the aspect of model architecture, the DAIF-Net is supposed to take a focal stack with an arbitrary number of images and output the underlying AIF image and depth map. We modify the encoder of the U-Net \cite{ronneberger2015u}, and design our DAIF-Net, as shown in \Cref{fig:funet}. In order to support a flexible number of input branches, every image will be passed through the same encoder and bottleneck. 

In order to learn the depth and AIF image from a focal stack simultaneously, we observe that sharpness is the connection between them. Intuitively, a sharper point indicates a closer distance between the depth and the focus distance. Meanwhile, sharper points better preserve the corresponding colors of the AIF images. Therefore, training a model that is sensitive to the sharpest points in the focal stacks is beneficial to predicting both AIF images and depth maps. To measure the sharpness, our designed model exploits the difference between image pixel-wise local feature maps, $i.e.$, the absolute sharpness, and the stack-wise global feature maps, $i.e.$, the sharpest regions. To compare the local feature maps and the global feature maps, our model adopts layer-wise global pooling from DefocusNet \cite{9157368}. Specifically, the model computes the output of the convolution block for every input image to produce the local feature maps. Then the maximum values are selected from the local feature maps across all branches to generate the global feature maps. The global feature maps are then concatenated to the local feature maps and the combined features are passed into the next convolution block. This way, global and local feature maps are visible to every branch. By comparison, the sharper regions will be highlighted by the multi-input encoder. The decoder then produces the AIF image and the depth jointly from the sharpness clue. 

Note that the inputs to the model have 4 channels: the RGB color channels and the focus distance channel, where we expand the focus distance to the size of the image, and concatenate it to the color channels. The focus distances can be extracted from the EXIF properties, which come together with the image and can be acquired with minimal cost. The focus distance is an essential input in our model. It increases the absolute accuracy of depth estimation, while not affecting the self-supervised characteristics of the framework.

\subsection{Self-supervised Training} \label{sec:self_supervise}
While our DAIF-Net is trained by estimating an accurate depth map and AIF image that can perfectly reconstruct the input focal stack using the optical model, it is hard to achieve this goal solely using the reconstruction loss in practice. Noticing that the $CoC$ can have the same value in front of and behind the focus point, and the defocus is ambiguous in texture-less regions, we delicately design auxiliary losses and regularization to support our self-supervised training. First, we briefly introduce our reconstruction loss.

\textbf{Reconstruction Loss.} We adopt the reconstruction loss from \cite{8954440} and have:
\begin{equation}
    \mathcal{L}_{recon} = \frac{1}{N}\sum^N\alpha\frac{1-SSIM(\Bar{J}, J)}{2} + (1-\alpha)\|\Bar{J}-J\|_1,
\end{equation}
where $J$ is the input focal stack and $\Bar{J}$ is the reconstructed focal stack, \textit{SSIM} is the Structural Similarity measure \cite{1284395}, and $\alpha$ is used to balance the scale between the \textit{SSIM} and the $L_1$ loss which we set to be 0.85. 

\textbf{Coarse AIF Image.} We notice that the defocus map is a natural clue for multi-focus fusion. Given a stack of defocus maps, we acquire the index of the minimum defocus value for each pixel position and fuse these sharpest pixels to form the coarse AIF image. If the defocus maps are accurate, this is the sharpest image we can produce without extra calculation. Meanwhile, since the defocus map is produced from the depth map, a clearer coarse AIF image indicates a more precise map, and thus a more accurate depth map. Therefore, we regularize the blurriness to encourage the images to be clear, so that the quality of depth map can be improved.

\textbf{Blurriness Regularization.} To evaluate the sharpness of AIF image, we apply a Laplacian filter for the edge map and calculate its variance. By taking the negative log sum of the normalized variance, the blurry estimation loss \cite{9464709} encourages the image to be as clear as possible:
\begin{equation}
    \begin{aligned}
        \mathcal{L}_{blur} = -\frac{1}{N}\sum^N_{c=1}\beta\log\Big(\frac{\Sigma_i\Sigma_j(\nabla^2\Hat{I}(i, j))^2}{M-\mu^2}\Big),
    \end{aligned}    
\end{equation}
where $I$ is the image, $M$ is the number of pixels, $\mu$ is the mean pixels value, $\nabla^2$ is the Laplacian operator and $\beta$ is a scale factor which we set to be 0.01. We apply the blur estimation loss on the predicted and the coarse AIF images to improve their sharpness. 

\textbf{Smoothness Regularization.} To prevent the estimated depth from drastic disparity in homogeneous regions and increase the consistency between the estimated depth and AIF image, we apply the smoothness prior to regularize the predicted depth map. The edge-aware smoothness regularization is:
\begin{equation}
    \mathcal{L}_{smooth} = \frac{1}{N}\sum^N |\partial_x \Bar{D}|e^{-\beta\partial_x \Bar{I}} + |\partial_y \Bar{D}|e^{-\beta\partial_y \Bar{I}},
\end{equation}
where $\Bar{I}$ is the predicted AIF image, $\Bar{D}$ is the predicted depth map, and $\beta$ is the scale factor for the edge sensitivity which we set to be 2.5. The overall loss then becomes:
\begin{equation}
    \begin{aligned}
        \mathcal{L} = \mathcal{L}_{recon} + \mathcal{L}_{blur}(\Bar{I}) + \mathcal{L}_{blur}(\Hat{I}) + \lambda \mathcal{L}_{smooth},
    \end{aligned}
\end{equation}
where $\Hat{I}$ is the coarse AIF image, and $\lambda$ is used to control the importance of the the smooth regularization, which we set to be 0.5.

\section{Experiments} \label{sec:empirical} 
\input{tables/defocus}
\input{tables/nyuv2}
In this section, we show qualitative and quantitative results of our proposed framework on a synthetic dataset \cite{9157368}, a real dataset with synthetic defocus blur \cite{8954440, 9464709} and a real focal stack dataset \cite{7298972}.

\subsection{Configurations} \label{sec:conf}
The framework is implemented in Pytorch \cite{10.5555/3454287.3455008}. We train the model using the Adam optimizer \cite{Kingma2015AdamAM} and the cosine annealing scheduler with an initial learning rate of $5e-4$. The batch is set to 32 and the focal stack size is 5 if not explicitly specified. We train our model for 2000 epochs on a single NVIDIA V100 GPU for each experiment. 

\subsection{Datasets} \label{sec:data}

\textbf{Synthetic Dataset} DefocusNet Dataset \cite{9157368} is a synthetic dataset rendered using Blender. The dataset consists of random objects distributed in front of a wall. A virtual optical camera takes five images of the scene with varying focus distances and forms a focal stack. The original dataset is wildly used in supervised methods \cite{9157368, Wang-ICCV-2021, yang2022deep}. However, the focus distances of the focal stacks are overly concentrated, causing indistinguishable defocus blur. Therefore, to perform an experiment in a similar setting, we regenerate the dataset with a set of more distributed focus distances using the code provided by the dataset author. The new dataset consists of 1000 focal stacks, with f-number $N=1.2$, focal length $f=2.9mm$, and focal distance $F$ at 0.3$m$, 0.45$m$, 0.75$m$, 1.2$m$ and 1.8$m$. The maximum depth is $3 m$.

\textbf{Real Image with Synthetic Defocus Blur.} To acquire sufficient realistic defocus images for model training and evaluation, we render the focal stack datasets from RGB-D datasets using the thin-lens equation and PSF convolution layers as in \cref{sec:optical}. The RGB-D datasets we use is NYUv2 dataset \cite{Silberman:ECCV12}. NYUv2 dataset with synthetic defocus blur \cite{8954440, 9464709} is also commonly used in DFD tasks. The dataset consists of 1449 pairs of aligned RGB images and depth maps. We train our model on the 795 training pairs and evaluate on the 654 testing pairs. To render the focal stacks, we choose focal length $f=50mm$, f-number $N=8$ and focus distance $F$ at 1$m$, 1.5$m$, 2.5$m$, 4$m$ and 6$m$ The maximum depth is $10 m$.

\textbf{Real Focal Stack Dataset.} Mobile Depth \cite{7298972} is a real-world DFD dataset captured by a Samsung Galaxy S3 mobile phone. It consists of 11 aligned focal stacks with 14 to 33 images per stack. Since neither depth ground-truth nor camera parameters are provided, we only perform qualitative evaluation and comparison on this dataset with no further finetuning.

\begin{figure*} [!ht]
    \centering
    \includegraphics[width=.9\textwidth]{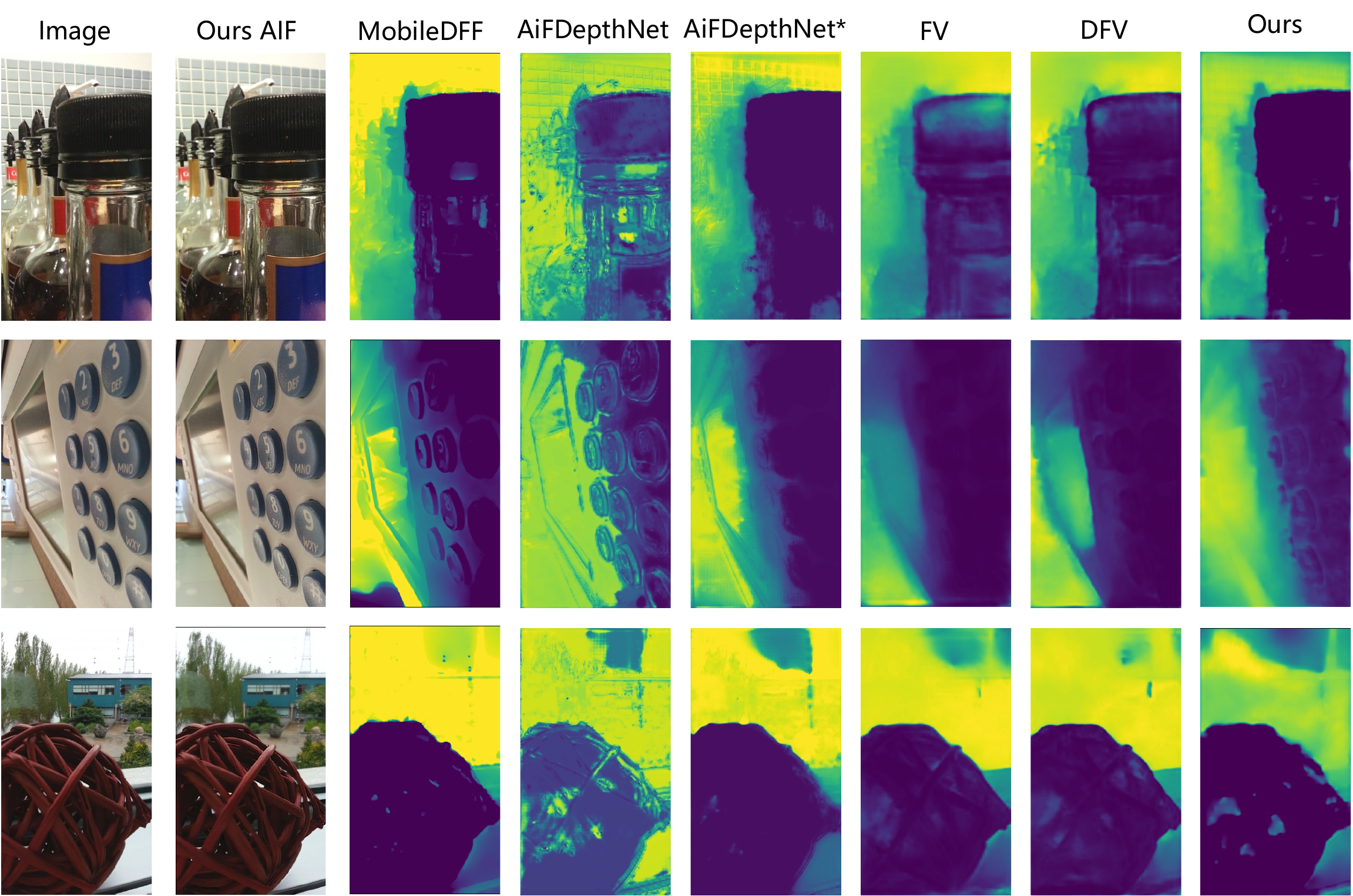}
    \caption{Qualitative depth estimation and AIF prediction results on the Mobile Depth dataset. The warmer color indicates a larger depth. Note that AiFDepthNet* is the finetuned model using AIF information.}
    \label{fig:mobiledff}
\end{figure*}

\subsection{Evaluation} \label{sec:eval}
\textbf{DefocusNet Dataset.} We split the 1000 focal stacks into 500 training focal stacks and 500 testing focal stacks. For a fair comparison, we trained our method, along with the open-source state-of-the-art DFD methods \cite{9157368, yang2022deep}, on our new training set. We present the qualitative results of the testing set in the supplementary material. The quantitative evaluation results are shown in \Cref{tab:defocus}. It is expected that our method does not perform as well as the supervised methods. This is because the DefocusNet dataset is texture-less, and our self-supervised framework is less sensitive to backgrounds, where defocus change is less obvious. Such issues do not exist for supervised methods because they always have access to the depth ground-truth. Meanwhile, we observe that our method is on par with the supervised methods when counting the results only for depths less than 0.5$m$, which indicates that our self-supervised method has higher accuracy in closer ranges.

\textbf{NYUv2 Dataset.} We present the results of our framework trained on the NYUv2 dataset with the synthetic focal stacks. In addition to Figure \ref{fig:head}, more qualitative results are provided in the supplementary material. We evaluate our models on the sparse testing focal stacks and compare our results to other DFD methods. \Cref{tab:nyuv2} shows that in sceneries with complex textures, our method is on par with the state-of-the-art on the majority of the metrics. Note that \cite{8954440, 9464709} rely on AIF image for training or inference; DFF-FV/DFV\cite{yang2022deep} are the state-of-the-art supervised DFD works on synthetic datasets. The result of AiFDepthNet \cite{Wang-ICCV-2021} is not presented because they have not released their training code. In contrast, our framework is completely self-supervised, with no AIF images or depth ground-truth required. 

\textbf{Mobile Depth Dataset.} To evaluate our model on real focal stacks, qualitative experiments are performed on the Mobile Depth dataset \cite{7298972}. We compare our model with MobileDFF \cite{7271087}, AiFDepthNet \cite{Wang-ICCV-2021}, DDF-DFV/FV \cite{yang2022deep} and the finetuned AiFDepthNet using AIF ground-truth. We also compare our generated AIF image with the AIF image produced by MobileDFF. Figure \ref{fig:mobiledff} shows parts of the results. Among these, MobileDFF is an optimization-based method and takes the full focal stack ranging from 14 to 30 images per stack. The resting deep methods generate depth using 5 uniformly sampled images from focal stacks. Since the dataset has no focus distance provided, we evenly assign the focus distances and normalize resulting depth map. From the figure, we can find that DFD models are generally sensitive to closer objects and edges. This is because the defocus change is more drastic at a close distance as demonstrated in Figure \ref{fig:defocus}, and image defocus is more obvious at the edge than in the regions without textures. Such characteristic is more apparent in our framework since our framework estimates depth completely depends on defocus information. While most of the deep methods give reasonable depth estimation, we claim that our framework is more advantageous since it is fully self-supervised.

\input{tables/ablation}

\section{Discussion}
In this section, we provide the ablation study on regularization, camera parameter selection and data generation process. We also discuss some of the limitations of our framework. Further discussion on the selection of hyperparameters and extension to the framework are provided in the supplementary material.
\subsection{Ablation Study} \label{sec:abs_study}

\textbf{Regularization Selection.} In our work, we propose that by regularizing the blurriness of the coarse AIF and the smoothness of the estimated depth, we can improve the model performance. As shown in \Cref{tab:abs}, we perform ablation studies using NYUv2 Dataset. From the table, we can observe that the regularization improves the accuracy ($\delta$1) of our model by $1.6\%$ and $3.8\%$ respectively. Our model also celebrates a drop in error rates (RMSE, AbsRel) with complete regularization. Therefore, the experiments prove the effectiveness of our regularization.

\textbf{Camera Parameters.} We perform an experiment to show that knowledge of the camera parameters is not necessary. Instead of using the camera parameter used in rendering the NYUv2 focal stack dataset, we train our model on another set of camera parameters: f-number $N = 1.2$, focal distance $f = 17mm$. The result is shown in the second last row of \Cref{tab:abs}. We can see that even being trained with different sets of camera parameters, our framework still provides promising results. Note that although we can train our model on different camera parameters, the camera parameters need to be selected so that it generates distinguishable defocus blurs as discussed in \cref{sec:defocus}.

\textbf{Data Generation Process.} To illustrate that the high performance of our method on the NYUv2 dataset is not because of the matching between the generated focal stack and the reconstructed defocus images, we slightly modify the data generation process. To be specific, we add Gaussian noise to every rendered defocus image and train our model on the noisy NYUv2 Focal Stack. The result is shown in the last row of \Cref{tab:abs}. We can see that the performance of our model does not vary even noise is added, which proves that our method is robust against noise and works regardless of the defocus generation process.

\subsection{Limitations}
Our DAIF-Net predicts the depth and AIF image by solving \cref{eq:reverse}. Theoretically, a focal stack with two images is enough to solve the equation. However, in practice, more images with distinct defocus blurs are needed. Intuitively, a larger focal stack helps the model to better find the sharpest point instead of the sharper point. In addition, two images may have close defocus values at a certain depth as shown in the right of \cref{fig:defocus}. Such indistinguishable defocus blurs make focal stack reconstruction and model optimization challenging. Therefore, we need more than two images in the focal stack so that distinguishable defocus values are available at every depth. To sum up, our framework is more suitable for closed scenes and textured scenes, where the defocus blurs are easier to be observed. 



\section{Conclusion}
In this paper, we first propose a real yet challenging setting for DFD tasks, where only sparse focal stacks are provided for training and testing. Then, to train a DFD model in such scenarios, we design a novel self-supervised framework for this task. In this framework, the DAIF-Net predicts depth and AIF image simultaneously from a focal stack. The model is then supervised by reconstructing the input focal stack using the optical model. Our framework relieves the need for the AIF or depth ground-truth, which makes its application to the real world more feasible. Experiments on various datasets also demonstrate the superior performance of our models, which set a strong baseline for the completely self-supervised DFD tasks. In the future, we plan to integrate the knowledge of the $CoC$ curve into the DAIF-Net. It is also necessary to establish a real focal stack dataset with focus distance and depth ground-truth included so that future DFD works can have a quantitative benchmark on more realistic data.

\noindent\textbf{Acknowledgements.} This research is supported in part by Shanghai AI Laboratory, the Natural Science Basic Research Program of Shaanxi under Grant 2021JQ-204, and the National Natural Science Foundation of China under Grant 62106183 and 62106182.

{\small
\bibliographystyle{ieee_fullname}
\bibliography{egbib}
}

\clearpage

\appendix
\section{Appendix}
\subsection{Synthetic Focal Stack Dataset for Robust Training}
As mentioned in Sec. 5.2, we generate a denser focal stack and sample five images in each iteration to train a more robust DAIF-Net model. To be specific, we render the dense focal stack datasets from NYUv2 dataset \cite{Silberman:ECCV12}. The dense focal stacks contain 100 images whose focus distance is uniformly distributed from 1$m$ to 9$m$. In the training stage, the dense focal stack is separated into 5 bins and one image is randomly sampled from each bin to form the sparse input focal stack. Therefore, our model benefits from the robustness of various focus distances. The same rendering strategy is applied to the evaluation set for best-model selection after each training epoch.

\begin{figure*}[]
    \centering
    \includegraphics[trim=165 715 180 760,clip,width=\textwidth]{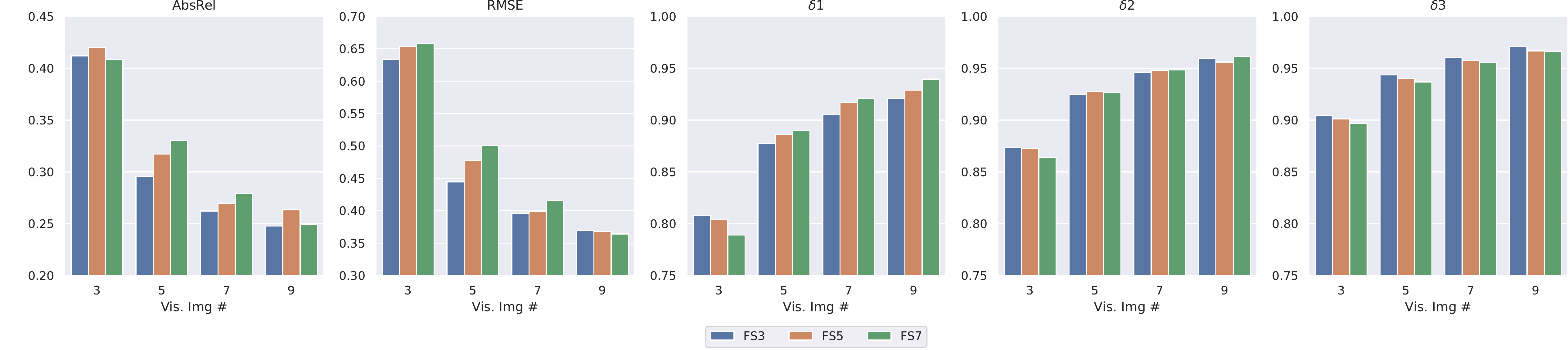}
    \caption{Robustly trained model generalizing to focal stack with different size and focus distances. Different colors indicate the different sizes of the training focal stack. Vis. Img \# indicates the size of the testing focal stack.}
    \label{fig:fs}
\end{figure*}

\subsection{Focus Distance Generalization} \label{sec:generalization}
Currently, all of the DFD works tend to train their models on datasets with fixed focus distances. In order to generalize our model to varying focus distances, we also propose a robust training strategy. When preparing the training data, we generate a denser focal stack, and at each iteration, we sample five images from the dense focal stack as the input. Since the focal stack is no longer a fixed input, the model can learn how different focus distances can affect the defocus better. To illustrate this point, we perform the robust training on the NYUv2 Dataset. The results, as shown in \Cref{fig:fs}, indicate that the robustly trained model, although having lower performance compared with the original models, can generalize better to focal stacks with different focus distances and with different numbers of images. 

\subsection{Ablation Study for Scales of the Losses}
Our ablation study focuses on the relative scale reconstruction loss and the blurriness loss for coarse AIFs. The results are shown in Table \ref{tab:rebuttalabl}. We can see from the table that the reconstruction loss contributes the most to the framework, while appropriate coarse AIF blurriness loss can further boost the model performance.

\subsection{Qualitative Results on DefocusNet Dataset}
We provide some qualitative results of our method compared with the state-of-the-art supervised works \cite{9157368, yang2022deep} on our rendered DefocusNet dataset \cite{9157368} in \Cref{fig:defocus}. Note our method is the only method that can predict AIF images together with the depth map. Also, we observe that our predicted depth map is on par with the supervised method at a closer distance.
\begin{figure*}[!h]
    \centering
    \includegraphics[trim=30 0 0 5,clip,width=0.9\textwidth]{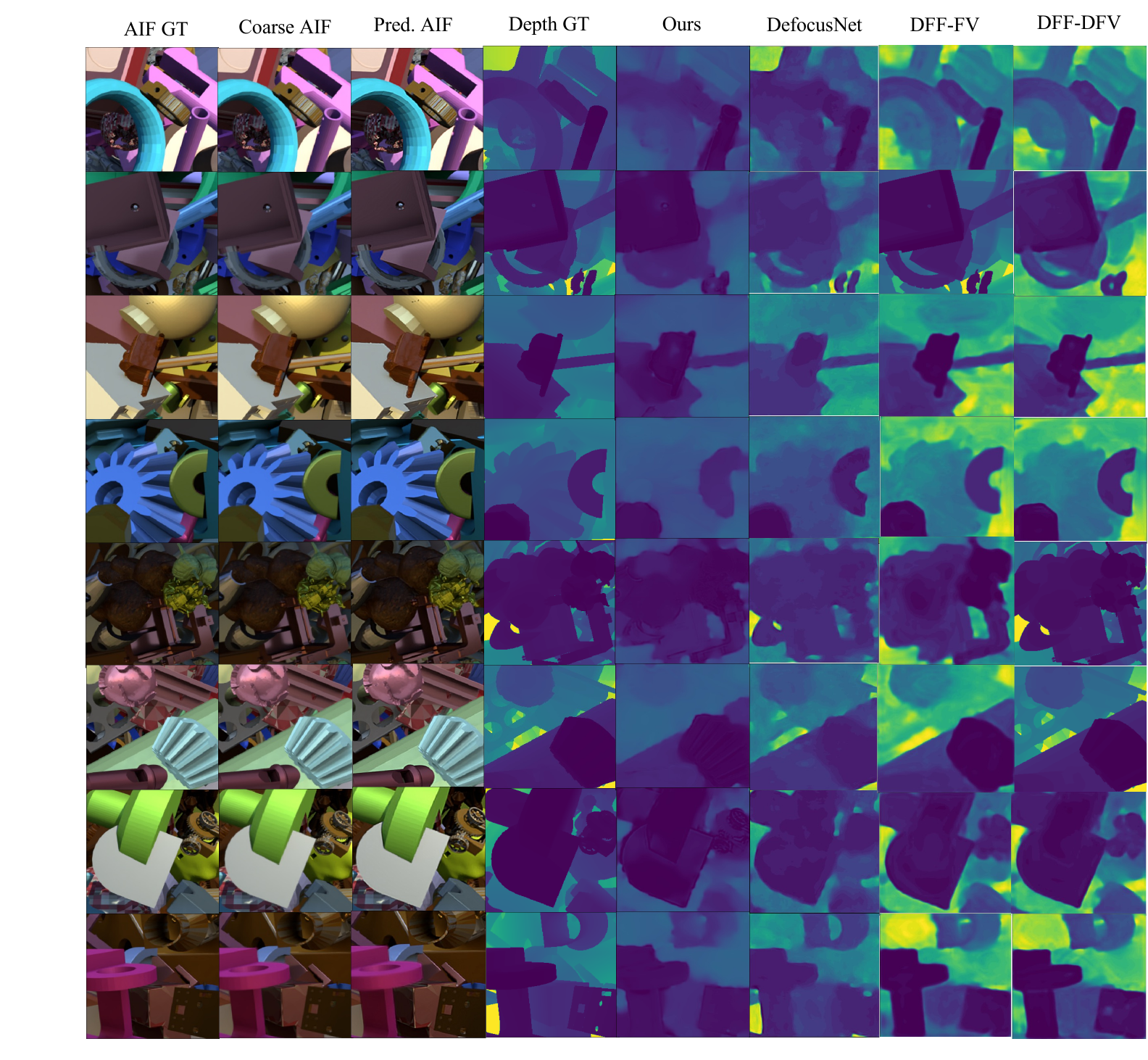}
    \caption{Some examples of the framework outputs comparing with the state-of-the-art supervised works. The outputs are produced from the input focal stacks with 5 images. For the depth map, lighter colors indicate farther distances.}
    \label{fig:defocus}
\end{figure*}

\subsection{Qualitative Results on NYUv2 Dataset}

\subsubsection{Predicted Depth Map and AIF images}
\begin{figure*} [!h]
    \centering
    \includegraphics[trim=60 40 50 35 ,clip,width=.9\textwidth]{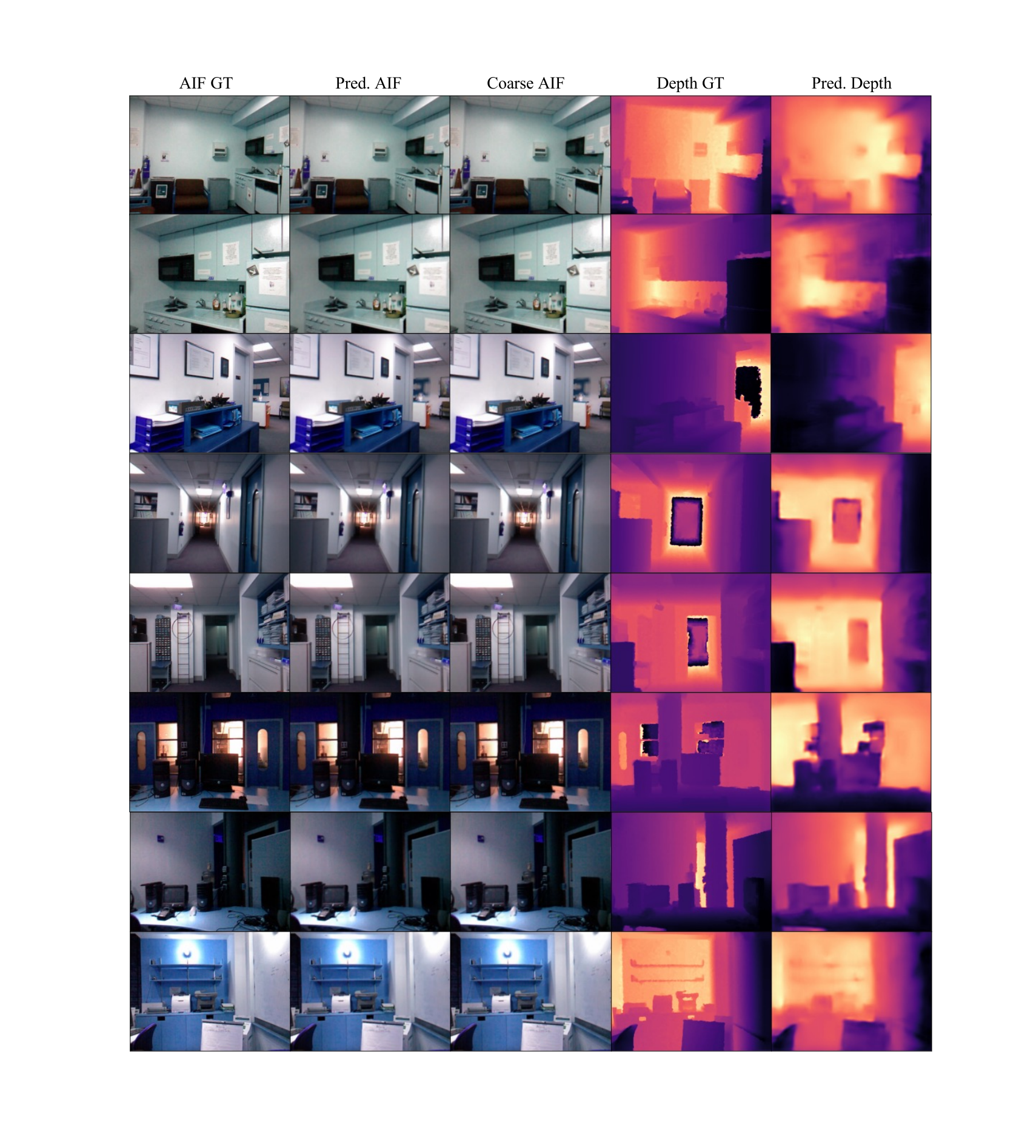}
    \caption{Some examples of the framework outputs. The outputs are produced from the input focal stacks with 5 images. For the depth map, lighter colors indicate farther distances.}
    \label{fig:aif_dpt}
\end{figure*}

Our DAIF-Net predict depth maps and the AIF images from the sparse focal stack. \Cref{fig:aif_dpt} shows some results of our framework. The results shows that our model performs well in the scenes with rich textures. Note we also present the coarse AIF images produced by picking the sharpest point in the focal stacks. The sharp coarse AIF images indicate a good quality of predicted depth maps.

\input{tables/abl_loss.tex}

\subsubsection{Defocus Map and Focal Stack}
\begin{figure*} [!h]
    \centering
    \includegraphics[trim=50 10 45 15 ,clip,width=.9\textwidth]{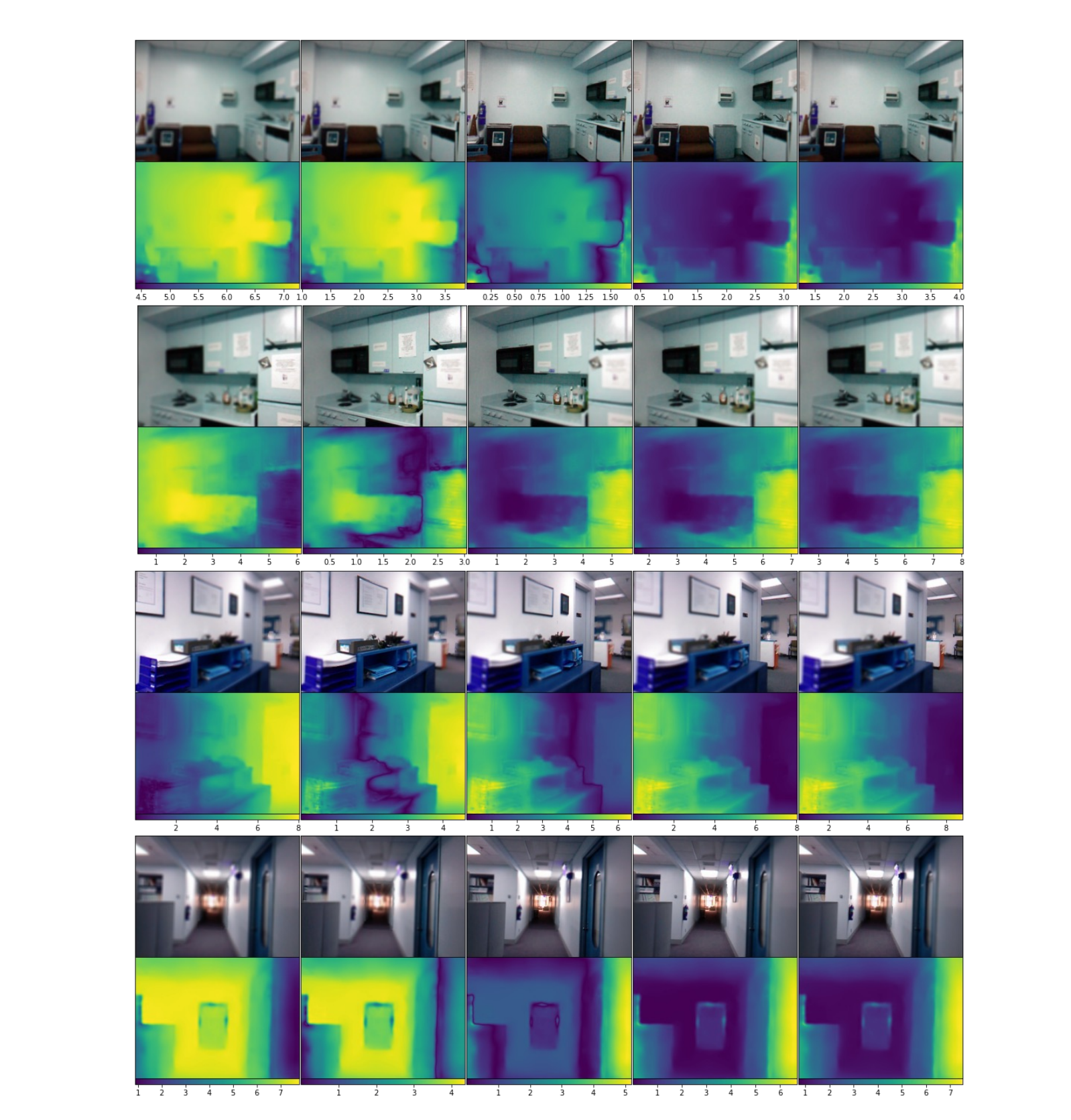}
    \caption{Some examples of the focal stack and their defocus maps. The focus distances are 1$m$, 1.5$m$, 2.5$m$, 4$m$ and 6$m$ from left to right. Darker colors indicate smaller defocus values.}
    \label{fig:df_fs}
\end{figure*}

We also provide the visualization of the defocus maps produced by the thin-lens module, and reconstruct the focal stack generarted by the PSF convolution layer. \Cref{fig:df_fs} shows some examples of the reconstructed focal stack along with their calculated defocus map. From the figure we can see that our optical model can reconstruct the focal stacks in a physical-realistic way, which is critical to the accurate prediction of the depth maps and AIF images.

\end{document}

%% file: tables/defocus.tex
\begin{table*}[!ht]
\centering
\begin{tabular}{ccccccccccc}
\hline
\multicolumn{1}{c|}{}                                              & \multicolumn{5}{c|}{Regular}                                                                                                & \multicolumn{5}{c}{0.5$m$}                                                                             \\ \cline{2-11} 
\multicolumn{1}{c|}{\multirow{-2}{*}{Methods}}                     & $\delta1 \uparrow$ & $\delta2 \uparrow$ & $\delta3 \uparrow$ & RMSE $\downarrow$ & \multicolumn{1}{c|}{AbsRel $\downarrow$} & $\delta1 \uparrow$ & $\delta2 \uparrow$ & $\delta3 \uparrow$ & RMSE $\downarrow$ & AbsRel $\downarrow$ \\ \hline
\multicolumn{11}{c}{\cellcolor[HTML]{EFEFEF}\textit{Supervised Learning}}                                                                                                                                                                                                                                 \\ \hline
\multicolumn{1}{c|}{DefocusNet \cite{9157368}}    & 0.912              & 0.967              & 0.983              & 0.194             & \multicolumn{1}{c|}{0.090}               & 0.911              & 0.933              & 0.938              & 0.062             & 0.069               \\
\multicolumn{1}{c|}{DFF-FV \cite{yang2022deep}}   & 0.883              & 0.953              & 0.980              & 0.231             & \multicolumn{1}{c|}{0.107}               & 0.977              & 0.996              & 0.999              & 0.023             & 0.032               \\
\multicolumn{1}{c|}{DFF-DFV  \cite{yang2022deep}} & 0.921              & 0.977              & 0.990              & 0.219             & \multicolumn{1}{c|}{0.104}               & 0.976              & 0.996              & 0.999              & 0.023             & 0.031               \\ \hline
\multicolumn{11}{c}{\cellcolor[HTML]{EFEFEF}\textit{Self-supervised Learning}}                                                                                                                                                                                                                            \\ \hline
\multicolumn{1}{c|}{Ours}                                          & 0.746              & 0.883              & 0.938              & 0.351             & \multicolumn{1}{c|}{0.177}               & 0.889              & 0.987              & 0.992              & 0.072             & 0.138               \\ \hline
\end{tabular}
\caption{Evaluation results on DefocusNet test set. Regular means all results are considered; $<0.5 m$ only counts results for depth less than 0.5 meters. The results indicates that our model is on par with the supervised state-of-the-arts in a closer range.}
\label{tab:defocus}
\end{table*}

%% file: tables/nyuv2.tex
\begin{table*}[ht!]
\centering

\begin{tabular}{ccccccc}
\hline
\multicolumn{1}{c|}{Methods}                            & \multicolumn{1}{c|}{Input}       & $\delta1 \uparrow$ & $\delta2 \uparrow$ & \multicolumn{1}{c|}{$\delta3 \uparrow$} & RMSE $\downarrow$ & AbsRel $\downarrow$ \\ \hline
\multicolumn{7}{c}{\cellcolor[HTML]{EFEFEF}\textit{Analytical Methods}}                                                                                                                                                  \\ \hline
\multicolumn{1}{c|}{Moeller et al. \cite{7271087}}                     & \multicolumn{1}{c|}{focal stack} & 0.670              & 0.778              & \multicolumn{1}{c|}{0.912}              & 0.985             & 0.263               \\
\multicolumn{1}{c|}{Suwajanakorn, Hernandez, and Seitz \cite{7298972}} & \multicolumn{1}{c|}{focal stack} & 0.688              & 0.802              & \multicolumn{1}{c|}{0.917}              & 0.950             & 0.250               \\ \hline
\multicolumn{7}{c}{\cellcolor[HTML]{EFEFEF}\textit{Self-sup w/ AIF}}                                                                                                                                                     \\ \hline
\multicolumn{1}{c|}{Gur and Wolf \cite{8954440}}                       & \multicolumn{1}{c|}{in-focus}    & 0.720              & 0.887              & \multicolumn{1}{c|}{0.951}              & 0.649             & 0.184               \\
\multicolumn{1}{c|}{Defocus-Net \cite{9464709}}                        & \multicolumn{1}{c|}{defocus}     & 0.732              & 0.887              & \multicolumn{1}{c|}{0.951}              & 0.623             & 0.176               \\
\multicolumn{1}{c|}{Focus-Net \cite{9464709}}                          & \multicolumn{1}{c|}{focal stack} & 0.748              & 0.892              & \multicolumn{1}{c|}{0.949}              & 0.611             & 0.172               \\ \hline
\multicolumn{7}{c}{\cellcolor[HTML]{EFEFEF}\textit{Supervised Learning}}                                                                                                                                                 \\ \hline
\multicolumn{1}{c|}{DFF-FV \cite{yang2022deep}}                             & \multicolumn{1}{c|}{focal stack} & 0.956              & 0.979              & \multicolumn{1}{c|}{0.988}              & 0.285             & 0.470               \\
\multicolumn{1}{c|}{DFF-DFV \cite{yang2022deep}}                            & \multicolumn{1}{c|}{focal stack} & 0.967              & 0.980              & \multicolumn{1}{c|}{0.990}              & 0.232             & 0.445               \\ \hline
\multicolumn{7}{c}{\cellcolor[HTML]{EFEFEF}\textit{Fully Self-Supervised Learning}}                                                                                                                                      \\ \hline
\multicolumn{1}{c|}{Ours}                               & \multicolumn{1}{c|}{focal stack} & 0.950              & 0.979              & \multicolumn{1}{c|}{0.987}              & 0.325             & 0.170               \\ \hline
\end{tabular}

\caption{Evaluation results on NYUv2 test set. Note that the results of \cite{7271087, 7298972, 9464709, 8954440} are recorded from \cite{9464709}. \textit{Self-sup w/ AIF} means that the method is self-supervised but utilizes AIF ground-truth. Results show that our method is on par with the state-of-the-art on NYUv2 dataset for the depth-from-defocus task. }
\label{tab:nyuv2}
\end{table*}

%% file: tables/ablation.tex
\begin{table}[]
\resizebox{\columnwidth}{!}{
\begin{tabular}{c|ccc|cc}
\hline
Setting             & $\delta1 \uparrow$ & $\delta2 \uparrow$ & $\delta3 \uparrow$ & RMSE $\downarrow$ & AbsRel $\downarrow$\\ \hline
w/o cAIF blur & 0.934  & 0.973  & 0.986  & 0.462 & 0.254  \\ 
w/o smooth & 0.912  & 0.976  & 0.984 & 0.477 & 0.278 \\ \hline
Ours Original       & 0.950  & 0.979  & 0.987 & 0.325 & 0.170  \\ \hline 
Ours New Param.       & 0.957  & 0.979  & 0.986 & 0.348 & 0.178  \\
Ours Noisy          & 0.943    & 0.980  & 0.986 &  0.340 & 0.189 \\ 
\hline
\end{tabular}
}
\caption{Ablation studies for the regularization, the camera parameters and data generation process.}
\label{tab:abs}
\end{table}

%% file: tables/abl_loss.tex
\begin{table}[]
\centering
\resizebox{0.7\columnwidth}{!}{
\begin{tabular}{c|cccc}
\hline
 \multirow{2}{*}{cAIF $\mathcal{L}_{blur}$}         & \multicolumn{4}{c}{$\mathcal{L}_{recon}$}          \\ 
 & 1 & 10    & 100            & 1000  \\ \cline{2-5} 
0.1       & - & 0.904 & 0.934          & -     \\
1         & - & -     & -              & 0.917 \\
10        & - & -     & \textbf{0.950} & -    \\ \hline
\end{tabular}
}
\caption{The ablation study of loss scales. The reported numbers are the $\delta$ accuracies. - means the set of parameters fails.}
\label{tab:rebuttalabl}
\end{table}